\documentclass[aip,pop,reprint,a4paper]{revtex4-1}
\usepackage{cancel}
\usepackage{amsmath}
\usepackage{amssymb}
\usepackage{graphicx}% Include figure files
\usepackage{dcolumn}% Align table columns on decimal point
\usepackage{bm}% bold math
\usepackage{upgreek}
\usepackage{caption}
\usepackage{subcaption}%subfigure
\usepackage{comment}% make comments
\usepackage{color}% color your text

\newcommand{\xsf}{\bm{\mathsf{X}}}

\newcommand{\vectorTwo}[2]{\left(\begin{array}{c}#1 \\ #2 \end{array} \right)}

\newcommand{\matrixTwobyTwo}[4]{\left(\begin{array}{cc} #1 & #2 \\ #3 & #4 \end{array} \right)}

\includecomment{comment}     %show comments
\begin{document}

%\preprint{}

\title{Toroidal regularization of the guiding center Lagrangian} %Title of paper



\author{J. W. Burby}
 \affiliation{Courant Institute of Mathematical Sciences, New York, New York 10012, USA}
\author{C. L. Ellison}
 \affiliation{Lawrence Livermore National Laboratory, Livermore, CA 94550, USA}
%\thanks{}


\date{\today}

\begin{abstract}
In the Lagrangian theory of guiding center motion, an effective magnetic field $\bm{B}^* = \bm{B}+(m/e)v_\parallel\nabla \times  {\bm{b}}$ appears prominently in the equations of motion. Because the parallel component of this field can vanish, there is a range of parallel velocities where the Lagrangian guiding center equations of motion are either ill-defined or very badly behaved. Moreover, the velocity dependence of $\bm{B}^*$  greatly complicates the identification of canonical variables, and therefore the formulation of symplectic integrators for guiding center dynamics. This Letter introduces a simple coordinate transformation that alleviates both of these problems simultaneously. In the new coordinates, the Liouville volume element is equal to the toroidal cotravariant component of the magnetic field.  Consequently, the large-velocity singularity is completely eliminated. Moreover, passing from the new coordinate system to canonical coordinates is extremely simple, even if the magnetic field is devoid of flux surfaces. We demonstrate the utility of this approach to regularizing the guiding center Lagrangian by presenting a new and stable one-step  variational integrator for guiding centers moving in arbitrary time-dependent electromagnetic fields.
\end{abstract}


\maketitle %\maketitle must follow title, authors, abstract and \pacs


%\section{Introduction}
%\begin{comment}
%{\noindent \color{blue} If we're submitting to a non-plasma journal, i'd suggest a few sentences explaining that guiding center theory is well important to plasma physics -CLE}\\
%{\color{red} I'm thinking yes to POP Letters at this point. I still might add something here though. -JWB} 
%\end{comment}

Without loss of generality, the Lagrangian for both guiding centers and gyrocenters in time-dependent electromagnetic fields $(\bm{A},\varphi)$ may be written
\begin{align}
\ell&=(e\bm{A}(\bm{X},t)+mv_\parallel \bm{b}(\bm{X},t))\cdot\dot{\bm{X}}-H(\bm{X},v_\parallel,t),
%(\bm{X},v_\parallel,t)
%&\hspace{9em}-H(\bm{X},v_\parallel,t),
%(\bm{X},v_\parallel,\dot{\bm{X}},\dot{v}_\parallel)
\end{align}
where $H$ is either the guiding center or gyrocenter Hamiltonian, as appropriate. \cite{Littlejohn_1983, Cary_2009,Hahm_Lee_Brizard_1988,Brizard_2007} 
 The Hamiltonian can always be written as $H=e\varphi+K(\bm{X},v_\parallel;\bm{E},\bm{B})$, where the gyrocenter kinetic energy $K$ depends parametrically on the potentials only through the gauge-invariant $\bm{E}=-\partial_t\bm{A}-\nabla\varphi$ and $\bm{B}=\nabla\times\bm{A}$. (The parametric dependence is in general nonlinear and nonlocal; see Ref.\,\onlinecite{Burby_Tronci_ppcf_2016} for the leading-order nonlocal terms.) This Letter is concerned with addressing a pair of computational and theoretical challenges associated with $\ell$. 

The first challenge is theoretical in nature, and it concerns unphysical infinities that appear in the Euler-Lagrange equations, which read
\begin{subequations}
\label{eq:standard_gc}
\begin{align}
\dot{v}_\parallel&=\frac{e}{m}\frac{\bm{E}^*\cdot\bm{B^*}}{B_\parallel^*}\label{normal_a_par}\\
\dot{\bm{X}}&=\frac{\bm{B}^*}{B_\parallel^*}\frac{\partial H/\partial v_\parallel}{m}+\frac{\bm{E}^*\times\bm{b}}{B_\parallel^*}.\label{normal_drift}
\end{align}
\end{subequations}
Here the effective magnetic field is given by
\begin{align}
\bm{B}^*&=\bm{B}+\frac{m}{e}v_\parallel\nabla\times\bm{b}\\
B^*_\parallel&=\bm{b}\cdot\bm{B}^*=B+\frac{m}{e}v_\parallel\bm{b}\cdot\nabla\times\bm{b},
\end{align}
and the effective electric field is given by
\begin{align}
\bm{E}^*=\bm{E}-\frac{m}{e}v_\parallel \partial_t\bm{b}-\frac{1}{e}\nabla K.
\end{align}
Because the Hamiltonian $H$ is a smooth function of $\bm{X}$ and $v_\parallel$, these equations of motion become infinite whenever $B_\parallel^*=0$. While such infinities occur only at values of $v_\parallel$ that technically violate the guiding center and gyrocenter ordering assumptions, they nevertheless lead to vexing inconsistencies and complications in kinetic theories built on top of $\ell$. For instance, because the support of a Maxwellian distribution function contains all of phase space, gyrokinetic theory necessarily allows for a small number of particles to sample the problematic range of parallel velocities. 

The second challenge is related to the problem of developing symplectic integrators for guiding center and gyrocenter motion. While the dynamical equations for guiding centers and gyrocenters, \eqref{normal_a_par}-\eqref{normal_drift}, possess a Hamiltonian structure, they are not written in canonical Hamiltonian form. Thus, the Hamiltonian structure is non-canonical, and standard symplectic integration techniques cannot be applied.\cite{Karasozen_2004} 

Previous authors have addressed each of these challenges with either computational or theoretical applications in mind. On the theoretical side, Correa-Restrepo and Wimmel\cite{Restrepo_1985} proposed a method for regularizing the infinities associated with $B_\parallel^*$ based on multiplying the $v_\parallel \bm{b}$-term in $\ell$ by a specially-designed form factor. 
%At values of $v_\parallel$ comparable with the thermal speed, the form factor is equal to $1$, which implies that the resulting regularized guiding center equations of motion agree with \eqref{normal_a_par}-\eqref{normal_drift} in regions of phase space that are consistent with the guiding center ordering. At larger values of $v_\parallel$, the form factor acquires a non-trivial $v_\parallel$-dependence that ensures infinities are not encountered in the equations of motion.
 %There has also been theoretical work on the identification of canonical coordinates for guiding center dynamics. 
White and Zakharov\cite{White_2003} proposed a system of canonical coordinates for guiding centers that may be used if the magnetic field admits nested toroidal flux surfaces. Zhang \emph{et al.}\cite{Zhang_R_2014} offer an alternative approach to guiding center canonical coordinates that is able to handle arbitrary magnetic fields, but leads to a cumbersome relationship between $(\bm{X},v_\parallel)$ and the canonical coordinates. On the computational side, the pioneering work on the development of structure-preserving integrators for guiding center motion was done by Qin and Guan,\cite{Qin_2008} who developed a variational integrator that performs well in axisymmetric magnetic field configurations. Subsequently, Ellison\cite{Ellison_thesis} investigated the prospect of extending the Qin-Guan technique to allow for arbitrary magnetic fields, but found the existing integrators to be numerically unstable. The latter work went on to uncover an unknown and serious gap in variational integration theory, and was able to identify stable variational integrators for a broad (yet not fully general) class of non-axisymmetric magnetic fields. The most flexible currently-available structure-preserving integrators for guiding center dynamics are the canonical-coordinate-based integrator of Zhang \emph{et al.},\cite{Zhang_R_2014}, and a special subset of the projected variational integrators developed recently by Kraus,\cite{Kraus_2017} all of which are capable of handling arbitrary magnetic fields. While each of these integrators is symplectic, none of them are derived directly from a discrete variational principle. Therefore, stable variational integrators that preserve a symplectic form on the guiding center phase space have yet to be identified.

In this Letter, we will argue that it is possible to simplify, unify, and generalize much of the previous work on regularization and canonization of guiding center theory. We will show that by applying a simple coordinate change to $\ell$, all $B_\parallel^*$ singularities, and all difficulties associated with finding canonical coordinates, are eliminated simultaneously; the theory is \emph{toroidally-regularized}. In order to illustrate the power of this result, we will then construct a novel and simple one-step variational integrator for guiding centers moving in arbitrary time-dependent electromagnetic fields without flux surfaces.

%In the ``nice'' coordinate system that we will discuss, the $B_\parallel^*$ denominators in \eqref{normal_a_par}-\eqref{normal_drift} are replaced with the (contravariant) toroidal component of the magnetic field. Thus, we refer to the process of moving into the nice coordinate system as \emph{toroidal regularization.} While we will focus our discussion on guiding center theory, toroidal regularization is applicable to both guiding center and gyrocenter dynamics. Moreover, toroidal regularization is flexible enough to handle time-dependent electromagnetic fields that may or may not possess nested magnetic flux surfaces. 

%For guiding centers moving in a time-independent magnetic field and zero electric field, the Hamiltonian is
%\begin{align}
%H=\frac{1}{2}m v_\parallel^2+\mu B,
%\end{align}
%where the parameter $\mu$ represents the magnetic moment. For the time being, I will focus the discussion on this particular Hamiltonian and set of electromagnetic fields. 

Derivation of the toroidally-regularized guiding center Lagrangian comprises two coordinate transformations. The first transformation is near-identity. The second transformation is non-perturbative, but extremely simple. The only
required assumptions are (a) that the standard guiding center ordering parameter satisfies $\epsilon=\rho/L\ll 1$ and (b) that the guiding centers of interest move in a toroidal region with a toroidal angle $\phi$ that satisfies
\begin{align}
\label{eq:nonzero_bphi}
|\bm{B}\cdot\nabla\phi|\equiv |B^\phi|>0.
\end{align}
Neither flux surfaces nor time-independence of the fields need to be assumed. The second assumption (b) is generally valid in the interior of devices envisioned for achieving magnetic confinement fusion, in particular in tokamaks and stellarators.\footnote{Poloidal regularization is possible in reversed field pinch configurations, but is theoretically more cumbersome because the poloidal angle is not well-defined at the magnetic axis.
}

The near-identity transformation maps the coordinates $(\bm{X},v_\parallel)$ to the new coordinates $(\bm{\mathsf{X}},\mathsf{v}_\parallel)$ using the Lie transform
\begin{align}
(\bm{\mathsf{X}},\mathsf{v}_\parallel)=\exp(G)(\bm{X},v_\parallel),
\end{align}
where $G$ is a time-dependent undetermined $O(\epsilon)$ vector field on $(\bm{X},v_\parallel)$-space. The guiding center Lagrangian is transformed into
\begin{gather}
\bar{\ell}(\bm{\mathsf{X}},\mathsf{v}_\parallel,\dot{\bm{\mathsf{X}}},\dot{\mathsf{v}}_\parallel)=(e\bm{A}(\bm{\mathsf{X}})+m \mathsf{v}_\parallel \bm{b}(\bm{\mathsf{X}}))\cdot\dot{\bm{\mathsf{X}}}\nonumber\\
+e G^{\bm{X}}\times\bm{B}(\bm{\mathsf{X}})\cdot\dot{\bm{\mathsf{X}}}-(H(\bm{\mathsf{X}},\mathsf{v}_\parallel)+e G^{\bm{X}}\cdot\bm{E})+O(\epsilon).
\end{gather}
The compontents of $G$ are chosen according to 
\begin{align}
G^{v_\parallel}&=0\\
G^{\bm{X}}&=-\frac{m}{e B^\phi} \mathsf{v}_\parallel\nabla\phi\times\bm{b}.
\end{align}
Because the unit vector along the magnetic field may be expressed (without approximation) as
\begin{align}
\bm{b}&=\frac{B}{B^\phi} \nabla\phi-\frac{B}{B^\phi}\bm{b}\times(\nabla\phi\times\bm{b}),
%R_o\nabla\phi+(\bm{b}-R_o\nabla\phi)\nonumber\\
%&=R_o\nabla\phi+\bigg(\left(\frac{B}{B^\phi}-R_o\right)\nabla\phi-\frac{B}{B^\phi}\bm{b}\times(\nabla\phi\times\bm{b})\bigg)\nonumber\\
%&=\frac{B}{R_oB^\phi} R_o\nabla\phi-\frac{B}{B^\phi}\bm{b}\times(\nabla\phi\times\bm{b}).
\end{align}
the transformed guiding center Lagrangian becomes
\begin{gather}
\bar{\ell}(\bm{\mathsf{X}},\mathsf{v}_\parallel,\dot{\bm{\mathsf{X}}},\dot{\mathsf{v}}_\parallel)=\left(e\bm{A}(\bm{\mathsf{X}})+m \mathsf{v}_\parallel \frac{B}{ B^\phi} \nabla\phi\right)\cdot\dot{\bm{\mathsf{X}}}\nonumber\\
-\left(H(\bm{\mathsf{X}},\mathsf{v}_\parallel)-m\mathsf{v}_\parallel\bm{b}\cdot \frac{\bm{E}\times\nabla\phi}{B^\phi}\right)+O(\epsilon).\label{lie_tran_lag}
\end{gather}
Note that the explicit form of the near-identity transformation is
\begin{align}
\bm{\mathsf{X}}&=\bm{X}-\frac{m}{e B^\phi(\bm{X})}v_\parallel(\nabla\phi\times\bm{b})(\bm{X})+O(\epsilon^2)\\
\mathsf{v}_\parallel&=v_\parallel.
\end{align}
Apparently this transformation amounts to a $v_\parallel$-dependent modification of the usual gyroradius vector.
%It is also worth mentioning that the higher-order terms in the expansion may be computed without difficulty. By using additional Lie transforms, it is always possible to ensure that the ``symplectic part" of the Lagrangian is given as it is in Eq.\,\eqref{lie_tran_lag} to all orders. 

The non-perturbative transformation maps the coordinates $(\bm{\mathsf{X}},\mathsf{v}_\parallel)$ to the coordinates $(\bm{\mathsf{X}},\mathsf{v}_\parallel^*)$ according to
\begin{align}\label{non_pert_trans}
\mathsf{v}_\parallel^*=\mathsf{v}_\parallel \frac{B}{R_o B^\phi},
\end{align}
where $R_0$ is an arbitrary constant with the dimensions of length. 
The Lagrangian finally becomes
\begin{align}
\ell^*(\bm{\mathsf{X}},\mathsf{v}_\parallel^*,\dot{\bm{\mathsf{X}}},\dot{\mathsf{v}}_\parallel^*)&=\left(e\bm{A}(\bm{\mathsf{X}})+m \mathsf{v}_\parallel^* R_o\nabla\phi\right)\cdot\dot{\bm{\mathsf{X}}}\nonumber\\
&\hspace{6em}-H^*(\bm{\mathsf{X}},\mathsf{v}_\parallel^*)+O(\epsilon),
\end{align}
where the new Hamiltonian is given by
\begin{align}
H^*(\bm{\mathsf{X}},\mathsf{v}_\parallel^*)&=e\varphi +K^*(\bm{\mathsf{X}},\mathsf{v}_\parallel^*)\\
K^*(\bm{\mathsf{X}},\mathsf{v}_\parallel^*)&=K(\bm{\mathsf{X}},\mathsf{v}_\parallel^*(R_o B^\phi/B))\nonumber\\
&\quad- m v_\parallel^* \bm{b}\cdot\frac{\bm{E}\times R_o\nabla\phi}{B}+O(\epsilon).
\end{align}
For guiding centers with $E\times B$ speed that is comparable to the thermal speed, the leading-order toroidally-regularized guiding center kinetic energy is given by
\begin{align}
K^*_{\text{gc}}(\bm{\mathsf{X}},\mathsf{v}_\parallel^*)=&\frac{1}{2}m\frac{(R_o B^\phi)^2}{B^2}\mathsf{v}_\parallel^{*2}+\mu B\nonumber\\
&-\frac{1}{2}m \frac{|\bm{E}_\perp|^2}{B^2}- m v_\parallel^* \bm{b}\cdot\frac{\bm{E}\times R_o\nabla\phi}{B}.
\end{align}

The Euler-Lagrange equations associated with the toroidally-regularized Lagrangian $\ell^*$ are given by
\begin{subequations}
\label{eq:regularized_gc}
\begin{align}
\dot{\mathsf{v}}_\parallel^*&=\frac{e}{m}\frac{\bm{B}\cdot \bm{E}^*}{R_o B^\phi}\\
\dot{\bm{\mathsf{X}}}&=\frac{\bm{B}}{R_o B^\phi}\frac{\partial H^*/\partial \mathsf{v}_\parallel^*}{m}+\frac{\bm{E}^*\times R_o\nabla\phi}{R_oB^\phi},
\end{align}
\end{subequations}
where the effective electric field is given by $\bm{E}^*=\bm{E}-e^{-1}\nabla K^*$. By the assumption \eqref{eq:nonzero_bphi}, these equtions of motion are free of singularities. We have therefore succeeded in eliminating the infinities present in the standard variational guiding center equations of motion. Moreover, the $v_\parallel$-dependent $B_\parallel^*$ denominators have been replaced with $\mathsf{v}_\parallel^*$-\emph{independent} denominators $R_o B^\phi$. This significantly simplifies the process of computing the current density generated by a distribution of gyrocenters in variational gyrokinetics and drift kinetics. In contrast, the regularization proposed in Ref.\,\onlinecite{Restrepo_1985} retains $v_\parallel$ dependence in the denominators. It is also instructive to compare what we have done here with Ref.\,\onlinecite{Scott_2017}, specifically the discussion surrounding Eqs.\,(42) and (94) therein. There it is explained that in a low beta and large aspect ratio tokamak it is justifiable to replace $\bm{b}$ with $R\nabla\phi$. This enables one to introduce a transformation akin to \eqref{non_pert_trans} that eliminates $v_\parallel$-dependent denominators from the gyrocenter equations of motion. Therefore toroidal regularization may be viewed as a generalization of the ideas in Ref.\,\onlinecite{Scott_2017} that allows for high-beta, arbitrary aspect ratio, fully-three-dimensional field configurations.

\begin{figure}
  \includegraphics[width=0.45\textwidth]{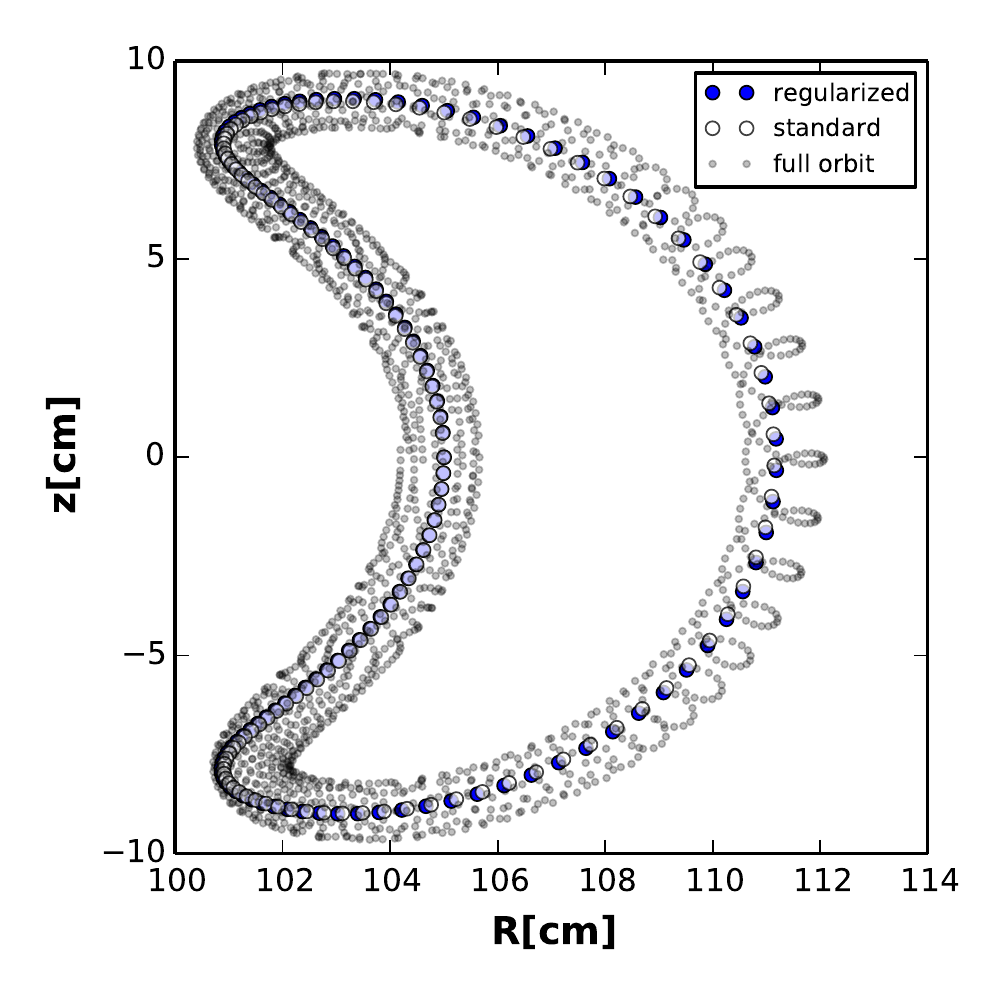}
  \caption{The toroidally-regularized guiding center equations accurately recover the familiar trapped particle ``banana orbit". Conditions: 2 keV proton, $B_0 = 1 $T, $R_0 = 100$cm, $q_0 = \sqrt{2}$, $(\bm{\mathsf{X}}, v_\parallel)$ = (5 cm, 0, 0, -12.9 cm/$\mu$s), $h= 0.3 \mu$s (100x smaller for full orbit).}
  \label{fig:banana_comparison}
\end{figure}

To verify the proposed transformation, Fig.~\ref{fig:banana_comparison} demonstrates that the regularized Lagrangian recovers familiar guiding center dynamics. We solve both the standard guiding center equations (Eq.\eqref{eq:standard_gc}) and the regularized guiding center equations (Eq.~\eqref{eq:regularized_gc}) for a trapped particle in an axisymmetric tokamak magnetic field. We use a system of toroidal coordinates $(r, \theta, \phi)$ and a magnetic field defined by the vector potential: \cite{Qin_2009}
\begin{align}
  \label{eq:axisymmetric_vector_potential}
  \bm{A}(r, \theta, \phi) =& \frac{B_0 R_0}{\cos^2 \theta} \left( r \cos \theta - R_0 \log\left(1 + \frac{r \cos \theta}{R_0}{}\right) \right) \nabla \theta\nonumber\\
 &\hspace{11em}- \frac{B_0 r^2}{2 q_0} \nabla \phi,
\end{align}
where $B_0$ is a magnetic field amplitude and $R_0$ is the major radius. In this demonstration, both of the guiding center theories accurately represent the gyro-averaged particle motion.

\begin{figure}
  % \centering
  % \begin{subfigure}[t]{0.5\textwidth}
  %   \includegraphics[width=\textwidth]{vectorfield_comparison.pdf}
  %   \caption{The $\theta$-component of the guiding center vector fields}
  %   \label{fig:singular_vectorfield}
  % \end{subfigure}%
  % ~
%  \begin{subfigure}[t]{0.45\textwidth}
    \includegraphics[width=0.45\textwidth]{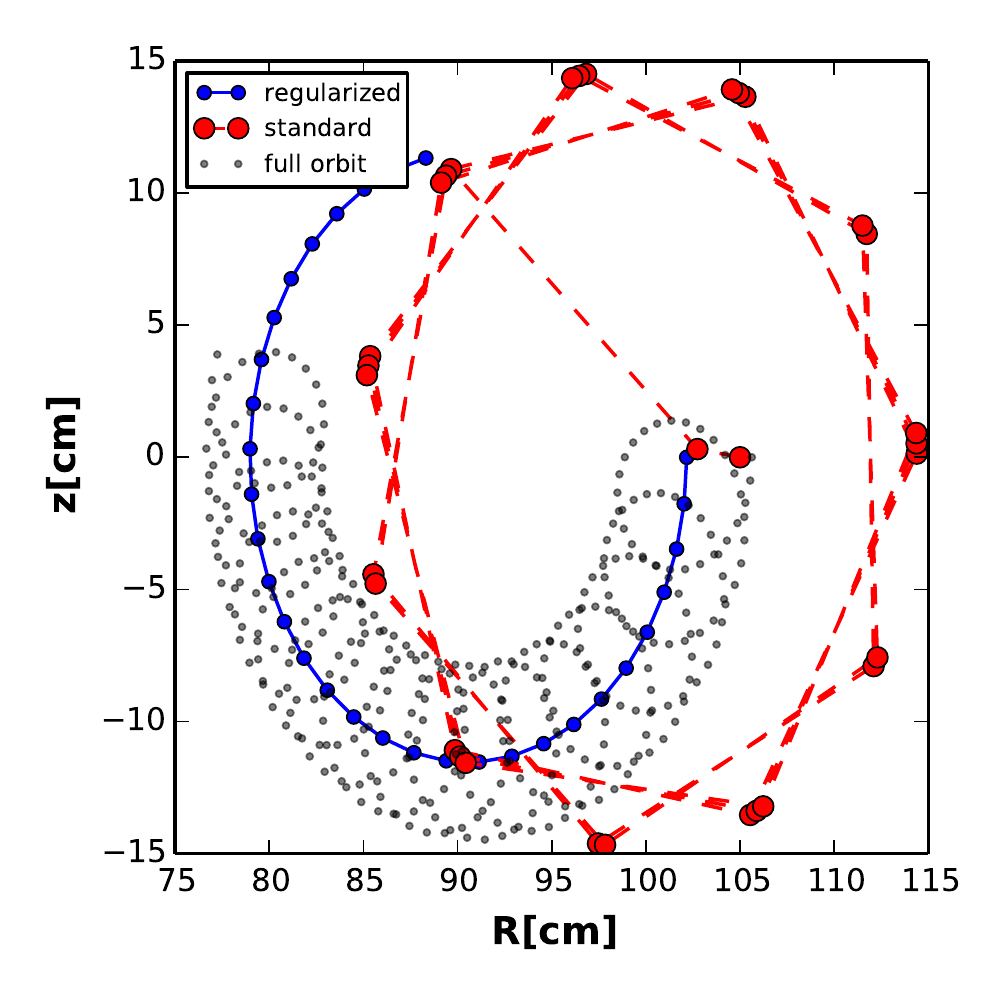}
 %   \caption{Guiding center and full orbit trajectories}
%    \label{fig:singular_trajectory}
%  \end{subfigure}
  \caption{The large-$v_\parallel$ singularity is manifest in a passing-particle trajectory generated by the standard guiding center equations. Same conditions as Fig.~\ref{fig:banana_comparison} except $q_0 = 0.1$, $v_\parallel=-600$cm/$\mu$s and timesteps reduced by a factor of ten.}
  \label{fig:singular_behavior}
\end{figure}

The successful elimination of the singular behavior is highlighted in Fig.~\ref{fig:singular_behavior}, where large-$v_\parallel$ trajectories are initialized near the $B_\parallel^*=0$ singularity. The toroidally-regularized Lagrangian produces a smooth trajectory that remains in good agreement with the full orbit calculation despite violating the guiding center ordering assumptions. Meanwhile, the trajectory generated by the conventional equations discontinuously leaps onto a different --- and more energetic --- trajectory upon encountering the singularity. 

Aside from the elimination of infinities, a significant benefit of toroidal regularization is that it greatly simplifies the identitification of canonical coordinates. Continuing with the $(r, \theta, \phi)$ toroidal coordinates (not necessarily field aligned) choose a gauge where $\bm{A}=A_r\nabla r+A_\theta\nabla\theta+A_\phi\nabla\phi$ satisfies $A_r=0$. In this gauge the Lagrangian $\ell^*$ becomes
\begin{align}
\label{eq:l_gauge}
\ell^*(\bm{\mathsf{X}},\mathsf{v}_\parallel^*,\dot{\bm{\mathsf{X}}},\dot{\mathsf{v}}_\parallel^*)=& e\bar{A}_\theta\dot{\theta}+(e\bar{A}_\phi+m\mathsf{v}_\parallel^* R_o)\dot{\phi}\nonumber\\
&\hspace{2em}-H^*(r,\phi,\theta,\mathsf{v}_\parallel^*)+O(\epsilon).
\end{align}
A viable set of canonical coordinates for guiding center motion, even in time-dependent fields, is therefore $(\theta,\phi,p_\theta,p_\phi)$, where
\begin{align}
p_\theta&=e \bar{A}_\theta\\
p_\phi&=e\bar{A}_\phi+m \mathsf{v}_\parallel^* R_o.\label{p_phi_eqn}
\end{align}
We have included the overbars here to emphasize the requirement of choosing the gauge $A_r=0$.
%\section{One-Step Variational Integrator}

Toroidal regularization also does not eliminate the usual benefits associated with Lagrangian guiding center theory. If the electromagnetic field is axisymmetric, and if the toroidal angle $\phi$ is assumed to be the symmetry angle, then the quantity $p_\phi$ in \eqref{p_phi_eqn} is conserved exactly. If the electromagnetic field is time-independent, the Hamiltonian $H=e\varphi+K$ is conserved exactly. Finally, phase space volume computed using the Liouville volume element 
\begin{align}
\Omega_L=e m R_0 B^\phi d^3\bm{\mathsf{X}}\,d\mathsf{v}_\parallel^*
\end{align}
is conserved. Note that $\Omega_L$ is free of $\mathsf{v}_\parallel^*$-depedendence.

Although symplectic integration of the regularized guiding center equations is facilitated by the preceding identification of canonical coordinates, it is simpler and computationally more efficient to directly advance the non-canonical coordinates. Toward that end, we turn now to the construction of a non-canonical symplectic integrator using the recently developed technique of ``degenerate variational integration.''\cite{Ellison_thesis, Ellison_2017} Degenerate variational integrators, or DVIs, were developed to remedy the numerical instabilities discovered in the initial (non-degenerate) variational guiding center integrators.\cite{Qin_2009, Li_2011, Ellison_2015_PPCF} Whereas it is only known how to construct a DVI for conventional guiding center dynamics under a restricted set of magnetic coordinates/magnetic field configurations, \cite{Ellison_thesis, Ellison_2017} the toroidally regularized Lagrangian is amenable to the method with no restrictions beyond the $A_r = 0$ gauge transformation.

To construct a DVI for the regularized system, begin with the Lagrangian in Eq.~\eqref{eq:l_gauge}. Further, for notational compactness, let $u = \mathsf{v}_\parallel^*$ and $\bm{A}^*(\xsf, u) = \bar{\bm{A}} + m u R_0 \nabla \phi$. Next, choose a discrete Lagrangian according to
\begin{align}
&  \ell_{\text{d}}(\xsf_{k}, u_k, \xsf_{k+1}, u_{k+1}) =  \nonumber\\
% = \ell^*(\xsf_{k+1}, u_{k+1}, \frac{\xsf_{k+1} - \xsf_k}{h}, \frac{u_{k+1} - u_k}{h} ) \nonumber \\
& \hspace{0.5em} e \bm{A}^*(\xsf_{k+1}, u_{k+1}) \cdot \frac{\xsf_{k+1} - \xsf_{k}}{h} - H(\xsf_{k+1}, u_{k+1}), 
\end{align}
where $h$ is the numerical step size. The discrete action corresponding to this choice is 
\begin{align}
  &S_d(\xsf_0, u_0, \xsf_1, u_1, ..., \xsf_N, u_N) = \nonumber\\
&\hspace{7em}\sum_{k=0}^{N-1} h \ell_d(\xsf_k, u_k, \xsf_{k+1}, u_{k+1}).
\end{align}
A variational integrator is obtained by requiring the variation of the discrete action with respect to each of the four coordinate functions $(\xsf, u)$ to be zero for all $k=1, ..., N-1$. The resulting \emph{discrete Euler-Lagrange equations} are given by:
\begin{subequations}
\label{eq:del}
\begin{align}
  &\nabla \bm{A}^*(\xsf_k, u_k) \cdot(\xsf_k - \xsf_{k-1}) - \bm{A}^*(\xsf_{k+1}, u_{k+1}) \nonumber\\
&\hspace{6em}+ \bm{A}^*(\xsf_k, u_k) - h \nabla H(\xsf_k, u_k) = 0 \label{eq:del_x} \\
  &\nabla_u \bm{A}^*(\xsf_k, u_k) \cdot(\xsf_k - \xsf_{k-1})\nonumber\\
&\hspace{11.5em} - h \nabla_u H(\xsf_k, u_k) = 0. \label{eq:del_u}
\end{align}
\end{subequations}
At first glance, this algorithm appears to be a multistep method, requiring specification of $\xsf, u$ at two instances in time before the time advance may be iterated. The crux of the DVI method is, however, that it avoids this multistep character. It is in fact possible to rearrange these equations into a one-step method as follows. Because $\bm{A}^*$ has only two non-zero components (namely, the $\theta$ and $\phi$ components), variables at time $t_{k+1}$ only appear in two components of Eq.~\eqref{eq:del}. The procedure for constructing a one-step method involves eliminating the $t_{k-1}$ dependence in these two equations. Let 
\begin{equation}
  \Delta = \vectorTwo{\Delta^\theta}{\Delta^\phi} = \vectorTwo{\theta_k - \theta_{k-1}}{\phi_k - \phi_{k-1}}.
\end{equation}
Then by Eq.~\eqref{eq:del}, $\Delta$ satisfies
\begin{equation}
  \matrixTwobyTwo{e A_{\theta,r}(\xsf_k)}{e A_{\phi,r}(\xsf_k)}{0}{m R_0} \vectorTwo{\Delta^\theta}{\Delta^\phi} = \vectorTwo{h H^*_{,r}(\xsf_k, u_k)}{h H^*_{,u}(\xsf_k, u_k)}.
\end{equation}
Pertinently, we can eliminate the $t_{k-1}$ dependence in Eq.~\eqref{eq:del} by expressing $\Delta$ as a function of the variables at time $t_k$. The one-step DVI, advancing $(\xsf_k, u_k)$ to $(\xsf_{k+1}, u_{k+1})$, is then given by:
\begin{subequations}
\begin{align}
   & e A_{\theta, r}(\xsf_{k+1})\left(\theta_{k+1} - \theta_{k} \right) + e A_{\phi, r}(\xsf_{k+1})\left(\phi_{k+1} - \phi_{k} \right) \nonumber\\
&\hspace{10em}- h H^*_{,r}(\xsf_{k+1}, u_{k+1})  = 0 \label{eq:onestepdel_r}\\
    &e A_{\theta, \theta}(\xsf_{k})\Delta^{\theta} + e A_{\phi, \theta}(\xsf_{k})\Delta^\phi + eA_\theta(\xsf_k)\nonumber\\
&\hspace{6em} - e A_\theta(\xsf_{k+1}) - h H^*_{,\theta}(\xsf_k, u_k)  = 0 \label{eq:onestepdel_theta} \\
    &e A_{\theta, \phi}(\xsf_{k})\Delta^\theta + e A_{\phi, \phi}(\xsf_{k})\Delta^\phi + 
eA^*_\phi(\xsf_k, u_k)\nonumber\\
&\hspace{4em} - e A^*_\phi(\xsf_{k+1}, u_{k+1}) - h H^*_{,\phi}(\xsf_k, u_k)  = 0 \label{eq:onestepdel_phi}\\
&mR_0 \left(\phi_{k+1} - \phi_{k} \right) - h H^*_{,u}(\xsf_{k+1}, u_{k+1})  = 0. \label{eq:onestepdel_u}
\end{align}
\end{subequations}
For time-dependent fields, the algorithm is unchanged except the field evaluations become, e.g., $A(\xsf_k) \mapsto A(\xsf_k, t_k)$. 

% Conservation properties
The DVI possesses, by construction, desirable conservation properties. For one, it can be observed in Eq.~\eqref{eq:onestepdel_phi} that the scheme exactly preserves the regularized version of the canonical toroidal momentum whenever toroidal symmetry is present. Additionally, the variational formulation of the algorithm implies that it preserves a symplectic two-form \cite{Marsden_2001} --- a fundamental property of Hamiltonian systems. Variational integrators constructed in this way preserve a two-form that is nearby to the one preserved by the continuous system, approaching it as the numerical step size tends to zero. \cite{Ellison_thesis, Ellison_2017} By preserving a symplectic two-form, the DVI retains the Hamiltonian character of the dynamics.

\begin{figure}[t]
  \centering
  \includegraphics[width=0.45\textwidth]{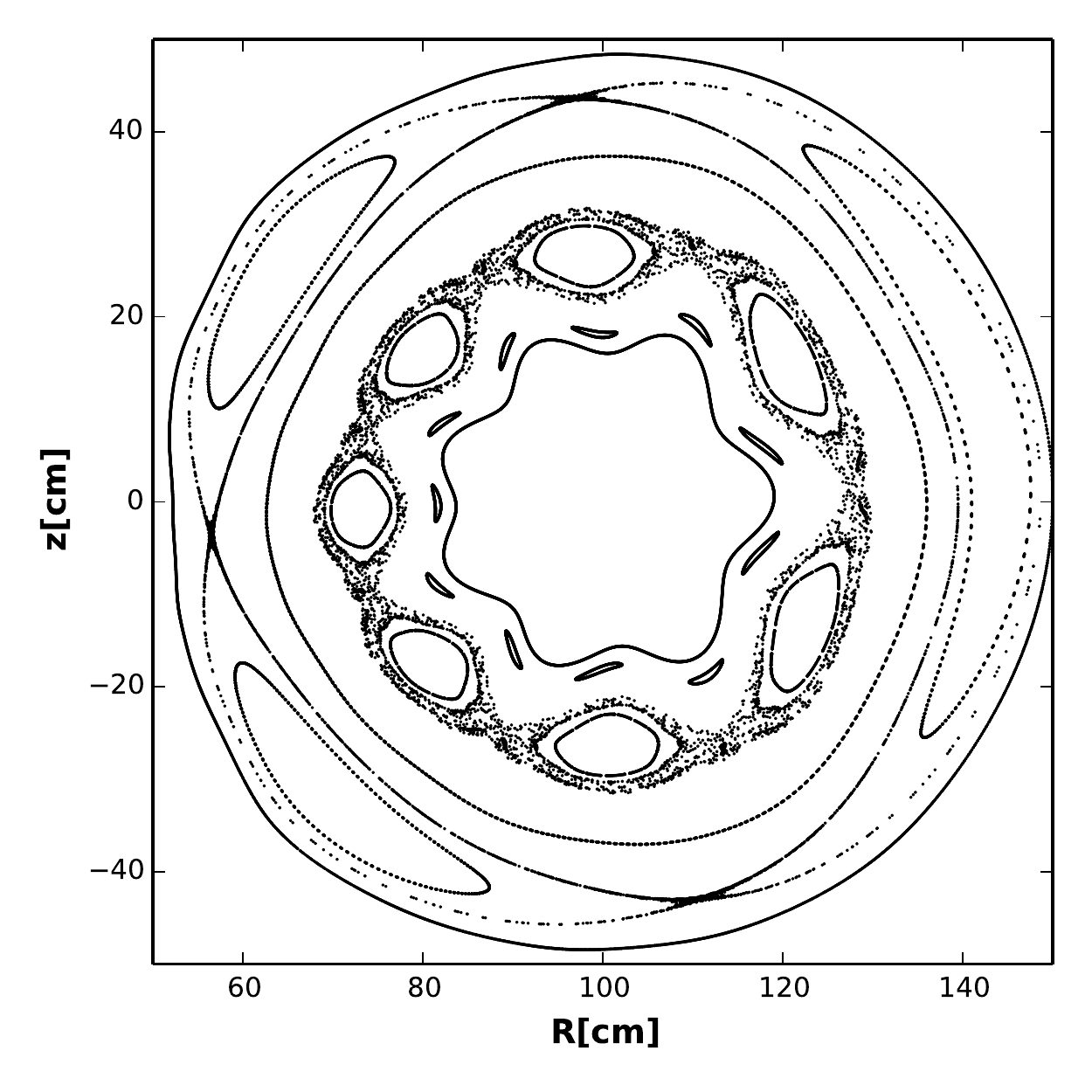}
  \caption{By retaining the Hamiltonian character of the dynamics, the symplectic DVI generates integrable and stochastic guiding center trajectories in the resonantly perturbed tokamak fields. Conditions: varying initial radii and poloidal angles; $q_0=1.35$; $\mathsf{v}_\parallel = 12.9$ cm/$\mu$s; zero magnetic moment; $h=3.5 \mu$s; $10^6$ steps taken.}
  \label{fig:poincare}
\end{figure}

% RMP Demo
To illustrate the benefits of non-canonical symplectic integration of guiding center trajectories, the final numerical study evolves passing particle trajectories in a resonantly perturbed tokamak. The resonantly perturbed field is described by the magnetic vector potential:
\begin{equation}
  \label{eq:rmp_vector_potential}
  \bm{A}(r, \theta, \phi) = \bm{A}_0 - \frac{B_0 r^2}{2 q_0} \sum_i \delta_i \sin(m_i \theta - n_i \phi) \nabla \phi,
\end{equation}
where $A_0$ is the axisymmetric vector potential given in Eq.~\eqref{eq:axisymmetric_vector_potential} and $\delta_i$ is the size of the $i$'th resonant perturbation. In this example we consider two perturbative harmonics: an $m=3, n=2$ harmonic and an $m=7, n=5$ harmonic, both of amplitude $\delta = 4 \times 10^{-4}$. Figure~\ref{fig:poincare} depicts a contant-energy Poincar\'e section formed by intersecting the particle trajectories with a plane of constant toroidal angle $\phi$. In the unperturbed, axisymmetric limit, the particle trajectories reside on circular KAM tori analogous to magnetic flux surfaces. Hamiltonian theory --- specifically, the KAM theorem --- dictates that these KAM tori should persist throughout a majority of the phase space when small perturbations are introduced. Because the DVI retains the Hamiltonian character, its Poincar\'e section can be seen to manifest this behavior for indefinitely long times. The same cannot be expected of non-symplectic algorithms, which eventually lose the Hamiltonian character of the dynamics to dissipative truncation error. 

To summarize, we have removed the unphysical infinities from variational guiding center theory in a simple and physically-appealing manner. As a result, we were able to find a very simple structure-preserving integrator for guiding centers. This integrator is capable of handling guiding center motion in arbitrary electric and magnetic fields, even those with time dependence and without nested magnetic flux surfaces. While we have not provided the explicit expressions here, analogous results apply in the context of variational gyrocenter motion. Thus, our results should enable the development of structure-preserving integrators for (at least) electrostatic drift kinetics and gyrokinetics. In addition, we have presented empirical evidence that our regularized guiding center theory performs surprisingly well at large parallel velocities, accurately capturing the shape (although not the phase) of the true orbit. Because such large parallel velocities violate the guiding center ordering assumption, it would be interesting to understand the reason for this good behavior in the future.

This research was supported by the U. S. Department of Energy, Office of Science, Fusion Energy Sciences under  Award No. DE-FG02-86ER53223  and the U.S. Department of Energy Fusion Energy Sciences Postdoctoral Research Program administered by the Oak Ridge Institute for Science and Education (ORISE) for the DOE. ORISE is managed by Oak Ridge Associated Universities (ORAU) under DOE contract number DE-AC05-06OR23100. All opinions expressed in this paper are the author's and do not necessarily reflect the policies and views of DOE, ORAU, or ORISE. This work was also performed under the auspices of the U.S. Department of Energy by Lawrence Livermore National Laboratory under contract DE-AC52-07NA27344. LLNL-JRNL-737871-DRAFT

%\bibliography{/Users/josh/Dropbox/Apps/Texpad/latex/cumulative_bib_file.bib}
\bibliography{SI_refs.bib}
%\bibliography{cumulative_bib_file_link.bib}
%%%%%%%%%%%%%%%%%%%%%%%%%%%%%%%%%%%%%

%%%%%%%%%%%%%%%%%%%%%%%%%%%%%%%%%%%%

\end{document}